\documentclass{egpubl}
\usepackage{eurovis2026}

\EuroVisPoster 

\usepackage[T1]{fontenc}
\usepackage{mathptmx} 
\usepackage[scaled=0.92]{helvet} 
\usepackage{courier} 

\usepackage{cite}  
\BibtexOrBiblatex
\electronicVersion
\PrintedOrElectronic
\ifpdf \usepackage[pdftex]{graphicx} \pdfcompresslevel=9
\else \usepackage[dvips]{graphicx} \fi

\usepackage{egweblnk}

\title[What Catches the Eye?]%
      {What Catches the Eye? A Conjoint Study of Infographic Design Preferences}

\author[A. K. Das  \& K. Mueller]
{\parbox{\textwidth}{\centering Amit Kumar Das\orcid{0000-0002-2600-8321}, Karanbir Pelia, Manav Nitesh Ukani, Klaus Mueller\orcid{0000-0002-0996-8590}
        }
        \\
{\parbox{\textwidth}{\centering Computer Science Department, Stony Brook University, USA
       }
}
}

\begin{document}

\maketitle

\begin{abstract}
Infographic designers balance many choices at once: chart type, color, and whether to add a benchmark or a scale. Past work studies these factors one at a time, so we know little about how readers weigh them against each other. We address this gap with a choice based conjoint study ($N = 65$) in which participants viewed pairs of infographics on a mock newspaper page about unemployment. Each infographic varied across three attributes: comparison type (none, US average, percentage scale), color (red, blue), and graphic type (single icon, icon series, bar chart). Comparison type drove most of the preference variation (58.5\%), followed by graphic type (29.2\%) and color (12.3\%). Readers favored percentage scale markers and benchmark comparisons; color had no practical effect. The percentage scale level adds axis information rather than a benchmark, so the comparison type result mixes two distinct ideas. A single topic and a narrow palette also limit external validity. We argue that conjoint analysis is a practical and underused tool for studying visualization preferences across many design dimensions.


\end{abstract}

\section{Introduction}

Infographic designers make many choices at once: chart type, color, and whether to add a benchmark or scale. Past visualization work studies these factors one at a time: memorability~\cite{Bor*13}, pictograph accuracy~\cite{HKF15}, and embellishments~\cite{Bat*10}. Single variable studies cannot show how readers weigh features against each other. Recent work models visualization preferences from large language models~\cite{Chu*24, DM26a,DTM26} but does not collect direct human judgments across multiple dimensions.

Conjoint analysis fills this gap~\cite{HHY14}. Participants pick between alternatives that vary across several attributes, and the method estimates how much each contributes to the choice. The technique is standard in marketing~\cite{GS90} and political science but has seen little use in visualization research.

We ran a conjoint study with 65 participants viewing pairs of infographics on a mock newspaper page about unemployment. Each infographic varied across three attributes: comparison type, color, and graphic type, yielding 18 conditions. We find that comparison type is the strongest driver of preference, graphic type matters less, and color barely matters. We argue that conjoint analysis is a practical method for visualization preference research.

\section{Study Design}

\begin{figure}[tb]
  \centering
  \includegraphics[width=0.88\columnwidth]{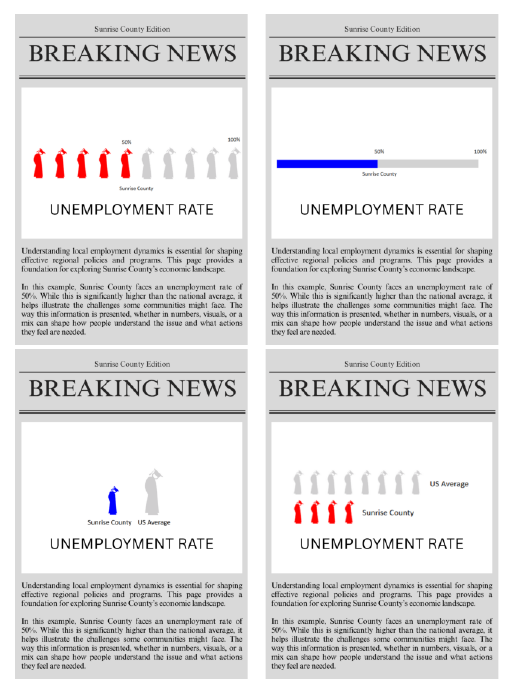}
  \caption{\label{fig:stimuli}
  Four of the 18 stimulus conditions, embedded in a mock newspaper page. The examples show different combinations of comparison type, color, and graphic type.}
\end{figure}

\textbf{Stimuli.}
We built a mock newspaper page titled ``Breaking News'' from a fictional Sunrise County Edition. The lead story reported a 50\% unemployment rate, with a short paragraph below the infographic giving neutral context. Each infographic varied across three attributes. \emph{Comparison type} had three levels: none (county rate only), US average (a labeled national value alongside the county rate), and percentage scale (markers at 50\% and 100\% on a visual axis). The percentage scale level adds calibration, not a benchmark; we return to this in the Discussion. \emph{Color} had two levels: red and blue. \emph{Graphic type} had three levels: a single person icon, a row of ISOTYPE style icons, and a horizontal bar chart. Crossing all levels gave $3 \times 2 \times 3 = 18$ conditions. Figure~\ref{fig:stimuli} shows four examples.

\textbf{Participants and procedure.}
We recruited 65 participants through Prolific. The largest groups came from the United States (25\%), the United Kingdom (15\%), and South Africa (10\%). Each participant completed about 12 paired comparison trials under a standard CBC protocol~\cite{HHY14}. Two newspaper pages appeared side by side, with left and right placement varied at random per trial. Participants picked the page they preferred and wrote a short justification of the choice. We delivered the study through Qualtrics and used Conjointly for design and analysis.

\textbf{Analysis.}
For each attribute level, the AMCE is the difference between the average choice probability when that level appears and the overall baseline, averaged across realizations of the other attributes~\cite{HHY14}. Relative importance is the range of preference share within each attribute divided by the sum of ranges, summing to 100\%. The effective sample is $65 \times 12 \times 2 = 1{,}560$ profile evaluations. The standard errors from Conjointly do not adjust for within respondent clustering; cluster robust corrections would inflate them by about 47\% at $\rho = 0.05$ but would not change the qualitative pattern. We therefore read magnitudes and patterns rather than report formal $p$ values.

\section{Results}

\begin{figure}[tb]
  \centering
  \includegraphics[width=0.88\columnwidth]{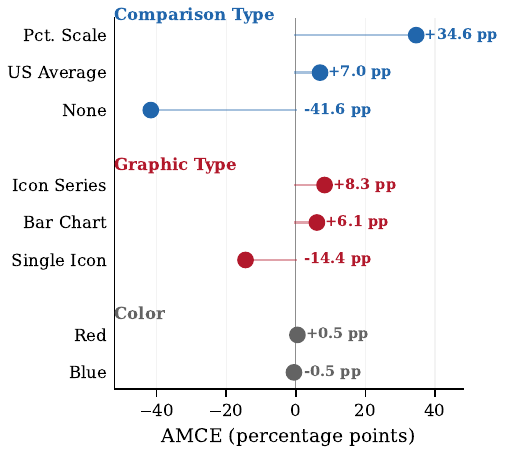}
  \caption{\label{fig:amce}
  Average marginal component effects (AMCEs). Positive values mean the level raises the chance of being chosen.}
\end{figure}

Comparison type accounted for 58.5\% of the preference variation, graphic type for 29.2\%, and color for 12.3\%. Figure~\ref{fig:amce} reports the AMCEs. The percentage scale level raised choice probability by 34.6~pp; the ``none'' level lowered it by 41.6~pp; the US average level raised it by 7.0~pp. Among graphic types, icon series and bar charts gained 8.3~pp and 6.1~pp, while single icons lost 14.4~pp. Red and blue differed by only 1~pp, with no practical effect.

\section{Discussion}

Readers value contextual reference more than aesthetic choices. This fits prior findings that overlays improve chart reading~\cite{KA12, DTM26b} and that comparison values aid risk understanding~\cite{GGG09}. Work on visualization trust~\cite{MPO25} treats clarity as the strongest driver, and Arunkumar et al.~\cite{APBB24} showed that axes and annotations shift perception toward ``information'' rather than ``image.'' The practical implication is to treat comparison and context as the first priority.

\textbf{The percentage scale confound.}
The high preference for the percentage scale condition needs care. The US average level adds a benchmark; the percentage scale level adds axis markers without a comparison value. One conveys a \emph{what}, the other a \emph{where}. This bundling makes the 58.5\% figure fragile. A more careful reading is that any contextual scaffolding beats no context (Appendix~\ref{app:factor_design}).

\textbf{Method sensitivity.}
Forced choice tasks can amplify the most salient attribute, since respondents adopt a lexicographic strategy~\cite{Koe*23}. Comparison features are the most visually distinct change in our stimuli, so their share is likely lifted relative to a single image rating. The dominance pattern is therefore method contingent: a rating based replication may yield a different attribute hierarchy (Appendix~\ref{app:method_sensitivity}).

\textbf{Graphic type and color.}
Readers preferred icon series and bar charts over single icons. Both formats give a visual sense of proportion, matching ISOTYPE findings~\cite{HKF15} and recent work on icon arrays for risk decisions~\cite{KLX25}. Color showed no effect. This null fits the affective visualization literature~\cite{LSZC21}: color preferences depend on topic and composition. Our controlled newspaper layout and unemployment topic may have neutralized differences between red and blue (Appendix~\ref{app:color_affect}).

\section{Conclusion}

Our conjoint study shows that comparison type drives most of the preference variation (58.5\%), followed by graphic type (29.2\%) and color (12.3\%). The leading percentage scale level adds axis information rather than a benchmark, so the result is best read as a preference for contextual scaffolding over no context. The single topic and narrow palette leave generalization open (Appendix~\ref{app:limitations}). Conjoint analysis fits multidimensional preference work in visualization, and its lightweight paired comparison format suits real evaluation flows. We release all stimuli, data, and analysis outputs as supplementary material.

\bibliographystyle{eg-alpha-doi}
\bibliography{references}

\clearpage
\appendix

\section{Limitations Overview}
\label{app:limitations}

We list the main limitations of the study and point to detailed treatments in the sections that follow:
\begin{itemize}
\item No cluster robust standard errors or bootstrap confidence intervals (Appendix~\ref{app:clustering}).
\item No covariates for numeracy, visualization literacy, or political orientation (Appendix~\ref{app:missing_covariates}).
\item Single topic and narrow color palette (Appendix~\ref{app:external_validity}).
\item Free text justifications collected but not analyzed (Appendix~\ref{app:quality}).
\item No interaction modeling across attributes (Appendix~\ref{app:interactions}).
\item No trial order or fatigue diagnostics (Appendix~\ref{app:trial_order}).
\item No formal power analysis before data collection (Appendix~\ref{app:power}).
\item Pair construction and dominated pairs not examined (Appendix~\ref{app:dominated_pairs}).
\item Relative importance metric reported without uncertainty (Appendix~\ref{app:importance_metric}).
\end{itemize}
Future work should adopt cluster robust inference, hierarchical Bayes or mixed logit models, and modern conjoint methods (Appendix~\ref{app:modern_conjoint}); include Bradley Terry scoring as a robustness check (Appendix~\ref{app:bradley_terry}); and triangulate preference with comprehension, trust, and image vs.\ information classification (Appendix~\ref{app:integrated_agenda}). Conjoint derived human preference weights could also serve as ground truth signals for computational preference models (Appendix~\ref{app:computational}).

\section{Participant Demographics}
\label{app:demographics}

Our sample of 65 participants was 55\% women and 45\% men. The largest age group was 25 to 34 years (55\%), followed by 45 to 54 years (25\%). The sample was well educated: 55\% held a bachelor's degree, 20\% a master's degree, 5\% a doctoral degree, and 10\% a professional degree. Participants came from seven countries, with the largest groups from the United States (25\%), the United Kingdom (15\%), and South Africa (10\%). We collected Big Five personality scores; the sample scored highest on conscientiousness ($M = 6.25$) and openness ($M = 5.53$). We did not model personality and preference interactions.

\section{Inference, Clustering, and Assumption Free Testing}
\label{app:clustering}

Our analysis relies on approximate standard errors derived from profile counts. Because each participant completed about 12 trials, the 1,560 profile evaluations are not independent. The standard correction is cluster robust (``sandwich'') standard errors:
\[
\widehat{\mathrm{Var}}_{\mathrm{CR}}(\hat{\beta}) = (X'X)^{-1} \left( \sum_{j=1}^{J} X_j' \hat{e}_j \hat{e}_j' X_j \right) (X'X)^{-1}
\]
where $J = 65$ clusters. The design effect inflation factor is $\sqrt{1 + (m-1)\rho}$; at $\rho = 0.05$ and $m = 24$ evaluations per respondent, SEs inflate by about 47\%. Even so, the comparison type effects ($+34.6$ and $-41.6$~pp) would likely remain significant, and color ($\pm 0.5$~pp) would remain null. With only 65 clusters the sandwich estimator can be biased; the \emph{wild cluster bootstrap} (Rademacher weights) and \emph{cluster jackknife} (leave one out) provide finite sample corrections, implemented in \texttt{sandwich} and \texttt{clubSandwich} in R.

\paragraph{Assumption free testing (CRT).}
\label{app:crt}
Ham et al.~\cite{HIJ22} proposed the conditional randomization test (CRT), which tests whether a factor matters \emph{in any way}: main effects, interactions, or nonlinear contributions. The test uses only the randomization of factor assignments and makes no distributional assumptions. The CRT sidesteps the clustering problem entirely and accepts any test statistic, including ML based ones, while providing exact finite sample $p$ values. In our setting it would test three nulls (comparison type, graphic type, color each irrelevant given the others). The open source \texttt{CRTConjoint} R package implements this for forced choice data.

\paragraph{Recommended re analysis protocol.}
Using the raw choice data in the supplementary material: (1) estimate AMCEs with cluster robust CIs via \texttt{cjoint::amce()}; (2) apply wild cluster bootstrap (10,000 reps) for refined CIs; (3) compute CRT $p$ values for each factor; (4) estimate Bradley Terry condition scores (Appendix~\ref{app:bradley_terry}) as a robustness check; (5) fit a linear probability model with comparison by graphic interaction and clustered SEs (Appendix~\ref{app:interactions}). Steps 1 to 5 require under 50 lines of R code. We intend to include at minimum steps (1) and (4), cluster robust AMCEs with 95\% CIs and a BT condition ranking, in the camera ready version of this paper, contingent on confirming that the Conjointly export provides the required choice level data format.

\section{Factor Design: Percentage Scale vs.\ Benchmark Comparison}
\label{app:factor_design}

The comparison type attribute bundles two qualitatively distinct manipulations. The ``US average'' level introduces an external benchmark; the ``percentage scale'' level introduces axis markers (50\% and 100\%) providing calibration without referencing an external value. One provides a \emph{what} (comparison value), the other a \emph{where} (position on a calibrated axis). This bundling makes relative importance fragile (the 58.5\% figure conflates benchmark and calibration effects), overstates the ``dominance'' claim (the two levels may appeal for different cognitive reasons), and should be resolved in future work via a 2 by 2 design crossing scale presence with benchmark presence. We retain the current analysis because the data cannot be re factored post hoc; readers should interpret ``comparison type'' as preference for \emph{any} contextual information over none.

\paragraph{Recommended reframing.}
Given this bundling, the headline claim is more precisely stated as: \emph{contextual scaffolding}, whether calibration (axis markers) or benchmark (reference value), is the strongest driver of preference, dominating graphic type and color. This ``contextual scaffolding $>$ none'' framing avoids implying that the two mechanisms are interchangeable and acknowledges that the 58.5\% relative importance figure reflects a composite of calibration and benchmark effects whose individual contributions remain unresolved.

\paragraph{Power target for a 2 by 2 follow up.}
Detecting a 10~pp main effect at $\alpha = 0.05$, 80\% power requires about 150 to 200 respondents (12 tasks each, $\rho \approx 0.05$). Detecting a scale by benchmark interaction of similar magnitude requires about 300 to 400 respondents. We recommend preregistering with a minimum of 200 (main effects) and aspirational 400 (interaction), using \texttt{cjoint} with cluster robust inference from the outset.

\section{Framing Text and Priming Concerns}
\label{app:framing}

The mock newspaper page originally stated unemployment was ``significantly higher than the national average,'' constituting a potential priming confound favoring the US average condition. The percentage scale condition is not directly primed; the ``none'' condition is most disadvantaged.

\paragraph{Sensitivity analysis.}
Assuming priming inflates the US average AMCE by $\delta$ pp:

\begin{center}
\small
\setlength{\tabcolsep}{3pt}
\begin{tabular}{lccc}
\hline
Scenario & US avg AMCE & Comp.\ RI & Pattern \\
\hline
Observed ($\delta=0$) & $+7.0$ pp & 58.5\% & Comp $>$ Gfx $>$ Col \\
Mild ($\delta=5$) & $+2.0$ pp & $\sim$52\% & Comp $>$ Gfx $>$ Col \\
Conservative ($\delta=10$) & $-3.0$ pp & $\sim$46\% & Comp $>$ Gfx $>$ Col \\
\hline
\end{tabular}
\end{center}

Even under the conservative scenario, comparison type retains the largest relative importance because the percentage scale AMCE ($+34.6$~pp) dominates. We recommend a neutral text replication; the supplementary materials include a revised paragraph.

\paragraph{Status of the priming text during data collection.}
We cannot confirm which version was used during collection, nor whether different participants may have seen different versions if the stimulus text was revised partway through data collection. The stimulus files in the supplement reflect the final state; reviewers can verify directly. If a mixed deployment occurred (some participants seeing the priming text and others the neutral version), the proportion per version could be estimated from the supplementary data if version metadata was logged, and AMCEs could be re estimated separately for each group to quantify the priming effect directly. We have no evidence that a mixed deployment occurred, but flag the possibility for completeness. The sensitivity analysis above bounds the threat regardless: even with 10~pp inflation, the overall finding holds.

\section{Visual Encoding of the US Average Condition}
\label{app:us_average}

In the US average condition, the national rate appeared as a labeled value alongside the county's rate. The exact numeric value, label text, font size, color, and proximity are documented in the stimulus image files (supplementary material); some figures may be hard to reconcile with the 50\% focal value without consulting those files directly.

The US average condition did not include axis markers (unlike the percentage scale condition), conveying comparison through juxtaposition rather than calibration. This lower visual salience may partly explain the preference gap via three mechanisms: (1) \emph{encoding density}: the percentage scale adds multiple elements (ticks, axis, labels) signaling ``more information''; (2) \emph{spatial organization}: axis markers impose a global framework that feels ``complete''; (3) \emph{familiarity}: axis based displays are standard, and familiarity increases both trust~\cite{MPO25} and preference. Future studies should systematically vary comparison information precision and salience (for example, label only vs.\ second glyph vs.\ calibrated axis).

\section{Sample, Covariates, and Subgroup Heterogeneity}
\label{app:sample}
\label{app:missing_covariates}
\label{app:heterogeneity}

Our 65 Prolific participants span seven countries (US 25\%, UK 15\%, SA 10\%). The sample is well educated (55\% bachelor's, 20\% master's) but not representative. We did not collect numeracy, visualization literacy, or political orientation, all plausible moderators.

\paragraph{Missing covariates.}
Numeracy moderates the effectiveness of graphical risk communication~\cite{GGG09}; high numeracy participants may show stronger preferences for comparison rich displays, inflating the comparison type effect. Visualization literacy affects graphic type evaluations. Political orientation interacts with our red and blue palette (see below). Recent predictive performance research~\cite{BLR26} demonstrates that user traits (including numeracy) and calibrated item difficulty jointly dominate interpretation accuracy, reinforcing the importance of collecting these covariates. Future studies should include the Berlin Numeracy Test, the VLAT, and political orientation measures to enable moderator analysis.

\paragraph{Subgroup heterogeneity.}
A descriptive split, US and UK (about 26 participants) vs.\ others (about 39), re estimating AMCEs with cluster robust SEs would check whether the comparison $>$ graphic $>$ color ordering is culturally consistent. With about 26 per subgroup, CIs will be wide; the goal is qualitative pattern checking, not significance testing.

\paragraph{Color and political associations.}
Red and blue is politically charged in the US (Republican and Democrat) but not identically elsewhere (UK: red = Labour). The aggregate null ($\pm 0.5$~pp) could mask offsetting effects. A descriptive US only ($n \approx 16$) vs.\ non US color AMCE comparison would be informative. We did not collect political orientation, so this hypothesis cannot be confirmed from the current data. Future studies should use a broader, politically neutral palette (for example, green, orange, teal).

\section{Data Quality, Free Text, and Choice Consistency}
\label{app:quality}
\label{app:dominated_pairs}

We applied no formal exclusion criteria beyond task completion. Participants provided free text justifications after each choice.

\paragraph{Pair construction and balancing.}
The Conjointly platform generates paired profiles using independent random assignment: on each trial, each attribute level is drawn uniformly and independently for both the left and right profiles. This ensures that, in expectation, all level combinations appear with equal frequency and that attribute levels are uncorrelated across profiles, the key assumption underlying unbiased AMCE estimation~\cite{HHY14}. The platform does not impose a D optimal or balanced incomplete block design; instead, it relies on the law of large numbers across trials and respondents to achieve approximate balance. With 65 respondents times about 12 trials, the design yields about 780 paired comparisons, and each of the 18 possible profiles appears as a left or right option about $780 \times 2 / 18 \approx 87$ times in aggregate, providing adequate coverage. However, specific pairwise matchups (of which there are $18 \times 17 / 2 = 153$ ordered pairs) are less densely covered (about 5 observations each), limiting the precision of condition level BT scores for rare pairings. The independent randomization also means that dominated pairs (where one profile is superior on all attributes) can arise by chance; we discuss their diagnostic value below.

\paragraph{Free text coding protocol.}
A light content coding pass would materially bolster the quantitative ordering: (1) \emph{keyword frequency}: code each response for mentions of scale or axis, comparison or benchmark, color, or graphic type; if comparison and scale mentions exceed 50\% and color falls under 10\%, this corroborates the conjoint ranking; (2) \emph{heuristic identification}: detect lexicographic strategies (always choosing the more informative option); (3) \emph{exclusion screening}: flag disengaged respondents (repeated or nonsensical text). A single coder applying binary tags to about 780 responses could complete this in a few hours.

\paragraph{Dominated pairs and choice consistency.}
Some randomly generated pairs may be \emph{dominated} (one profile superior on all attributes). Based on observed AMCEs, profiles combining percentage scale plus bar chart plausibly dominate those with no comparison plus single icon. The proportion of ``correct'' choices on such pairs serves as an attention or engagement diagnostic ($>$90\% suggests engaged responding). Near dominated pairs (superior on comparison plus graphic, inferior on color) should also produce consistent choices given color's negligible effect. We recommend future studies preregister a minimum 80\% dominated pair accuracy threshold as an exclusion criterion, and that any re analysis identify these pairs in the raw data.

\section{Interaction Effects}
\label{app:interactions}

Our analysis estimates only main effects (AMCEs). Plausible interactions include: (1) bar charts pairing naturally with percentage scales (both axis based); (2) icon arrays pairing with benchmarks (comparing icon sets); (3) color mattering more when comparison information is absent. With about 780 observations per three level attribute and no cluster robust inference, we have limited power for interaction tests (about 15~pp detectable; see Appendix~\ref{app:power}). Lexicographic processing~\cite{Koe*23} would attenuate interactions. A compact test would estimate a linear probability model with three preregistered interaction terms (bar by scale, icons by benchmark, red by none) and cluster robust SEs, feasible from the existing raw data.

\section{Relative Importance Metric}
\label{app:importance_metric}

The range share metric (range within attribute divided by sum of ranges) is sensitive to extreme levels. In our design, the ``none'' condition's $-41.6$~pp AMCE functions as an outlier, inflating the comparison type range to 76.2~pp. If importance were computed from the interquartile range or median absolute deviation, the comparison type advantage would narrow. The metric also lacks uncertainty quantification (no bootstrap CIs on the 58.5\%, 29.2\%, 12.3\% partition) and assumes additivity. AMCEs with respondent clustered CIs provide a more principled summary. Mixed logit or hierarchical Bayes estimation would further capture respondent level heterogeneity and produce distributions of importance weights rather than a single number.

\section{Method Sensitivity, Salience, and Rating}
\label{app:method_sensitivity}

Koesten et al.~\cite{Koe*23} found that forced choice tasks encourage lexicographic processing: participants attend to the most salient attribute first. In our stimuli, comparison type is the most visually distinctive change, likely amplifying its dominance relative to a single exposure evaluation. Color, while perceptually salient in isolation, does not change structural complexity, explaining its suppression.

\paragraph{Theoretical grounding.}
Fu and Li~\cite{FL25} formalized how attention and salience lead to amplified effect magnitude and possible \emph{importance reversals} between forced choice experiments and real world settings. Their model predicts that our comparison type AMCE is likely amplified relative to naturalistic single exposure viewing, and that the 58.5\% relative importance figure should not be generalized beyond side by side comparison.

\paragraph{Converging evidence from affordance elicitation protocols.}
Stokes et al.~\cite{Sto*25} systematically compared four methods for eliciting visualization affordances: free response, visualization ranking, conclusion ranking, and salience rating. They found that no single method fully replicates free response affordances. Crucially, ranking methods were influenced by participant bias toward certain chart types and by how suggested conclusions were framed, paralleling the salience amplification we observe in our CBC forced choice design. Their finding that GPT 4o performed best as a human proxy only for salience rating (but not for ranking or free response tasks) further underscores that method choice shapes which design attributes appear dominant. Our comparison type dominance should therefore be interpreted as method contingent: a salience rating or free response replication might yield a different attribute hierarchy.

\paragraph{Proposed rating based follow up.}
Each participant would view individual newspaper pages (not pairs) and rate each on a 7 point effectiveness scale. If comparison type still dominates, the finding is robust; if color and graphic type gain weight, salience explains part of the conjoint result.

\paragraph{Why no rating pilot was conducted.}
A within study rating arm would have materially strengthened the salience argument by directly demonstrating whether the forced choice format inflated the comparison type effect. We did not include one for two reasons: (1) the Conjointly platform is designed for CBC and does not natively support single stimulus rating tasks within the same survey flow, and (2) adding a separate rating block would have approximately doubled session length for a sample that was already modest ($N = 65$). Even a small between subjects pilot ($n \approx 20$ to $30$ rating, $n \approx 20$ to $30$ CBC) would have required separate recruitment and could not share stimuli seamlessly across platforms. We acknowledge this as a limitation and recommend that a future replication pair CBC and rating tasks using a unified survey tool.

\section{Color, Affect, and Context}
\label{app:color_affect}

Our null color result ($\pm 0.5$~pp) is consistent with the affective visualization literature. Lan et al.~\cite{LSZC21} found that color preferences interact with topic and composition. In our unemployment context, both red (urgency) and blue (trust) are plausible; neither carries a decisive advantage. A broader palette, including green, orange, and varying saturation, might reveal preferences our binary manipulation missed. The controlled newspaper layout may also have reduced color's visual weight. Lan et al.~\cite{LWC24} argued that affective dimensions are undervalued in visualization design; our null result does not contradict this but illustrates that detecting color effects requires wider gamuts, varied topics, or more sensitive measures (for example, affect ratings).

\paragraph{Affective infographic datasets.}
The InfoAffect dataset~\cite{FWHS26} provides 3,500 affect annotated real world infographics across six domains, revealing that affective responses depend heavily on the interplay between visual design and textual framing. Their finding that affect is content dependent supports our interpretation: color's null effect in an unemployment context does not imply a universal null, but rather that the affective valence of both red and blue is ambiguous for this topic. A follow up using InfoAffect's multi domain stimuli, or similarly varied topics, paired with explicit affect ratings alongside conjoint choice would test whether color gains importance when topics evoke clearer color and emotion associations (for example, green for environmental data, red for health emergencies).

\section{Computational Preferences and LLM Replication}
\label{app:computational}

DracoGPT~\cite{Chu*24} extracts visualization preferences from LLMs using the same paired comparison format we use. Our human data provide a benchmark for evaluating LLM derived preferences. Key divergence points: (1) LLMs lack perceptual grounding (our comparison type dominance may reflect visual salience that LLMs do not process); (2) LLMs may assign context independent color preferences that differ from our topic dependent null~\cite{LSZC21}; (3) multimodal LLM judges show position and verbosity biases~\cite{Che*24} absent in human responses.

\paragraph{Preference vs.\ quality assessment.}
It is important to distinguish our study's focus on \emph{preference} (which design do you choose?) from \emph{quality assessment} (how good is this design?). Emerging benchmarks such as VisJudge Bench~\cite{XVJ25} evaluate MLLMs as visualization quality judges across fidelity, expressiveness, and aesthetics, finding that even advanced models (for example, GPT 5) exhibit significant gaps with human expert ratings. Our conjoint data capture a complementary signal: not ``how good is this visualization'' but ``which design wins in a head to head comparison.'' This distinction matters because quality judgments involve absolute evaluation along multiple dimensions, whereas preference reveals relative trade off weights. Recent work on MLLM visualization literacy ~\cite{DM26a, DTM26, DTM26b} further shows that MLLMs and humans exhibit distinct failure patterns when interpreting charts: color palettes have no significant impact on MLLM accuracy, echoing our human null color finding, but plot type effects diverge from human patterns. Jointly, these results suggest that human conjoint preferences provide a ground truth signal that neither quality benchmarks nor MLLM judges can currently replace: our data reveal \emph{how humans weigh competing design attributes}, a signal orthogonal to single stimulus quality scores.

\paragraph{LLM replication protocol.}
Present 50 to 100 stimulus pairs to a multimodal LLM as side by side images, record binary choices across $k = 5$ to $10$ runs, estimate AMCEs, and compare with human results. If the attribute importance ordering matches, this suggests convergent validity; divergences identify dimensions needing human calibration. This requires only API access and can be completed in an afternoon.

\section{Complementary Outcomes and Integrated Design}
\label{app:trust}
\label{app:image_info}
\label{app:icon_bars}
\label{app:integrated_agenda}

Our study measures only preference; a complete evidence base should triangulate with additional outcomes.

\paragraph{Trust and clarity.}
McKinley et al.~\cite{MPO25} found clarity is the strongest driver of visualization trust, building on Pandey et al.'s~\cite{PMCO23} five dimensional framework. Our preferred conditions (percentage scale, US average) both increase informational clarity, suggesting that the preference for comparison type may partly reflect a trust response.

\paragraph{Cognitive affordances.}
Fygenson et al.~\cite{FPB25} proposed a framework of cognitive affordances in visualization that enumerates how design decisions (encoding, annotations, labels) and reader characteristics jointly determine a hierarchy of afforded information. Axis markers and benchmarks elevate task relevant affordances, calibration (``where is this value on the scale?'') and comparability (``how does this compare to a reference?''), to the top of the hierarchy. In a side by side forced choice task, these affordances become especially salient because participants are actively searching for discriminating features. The cognitive affordances lens thus provides a theoretical account for why comparison type features dominate: they increase the visualization's affordance for the specific cognitive action (comparison) that the task demands. This interpretation complements the trust and image vs.\ information accounts and predicts that the dominance of comparison type would attenuate in tasks that do not require comparative judgment.

\paragraph{Image vs.\ information.}
Arunkumar et al.~\cite{APBB24} found that axes and annotations shift perception toward ``information'' (easier to understand) versus ``image'' (more aesthetic). Our percentage scale and benchmark conditions would be classified as ``information,'' aligning with their finding that information classified visualizations elicit more positive judgments.

\paragraph{Annotations as preference drivers.}
Our findings connect to a broader pattern in the annotation literature: readers consistently prefer charts with more contextual scaffolding. Stokes et al.~\cite{SSC23} found that participants preferred heavily annotated line charts over minimally annotated or text only versions, and that annotation content and placement shaped both takeaways and preferences. Kong and Agrawala~\cite{KA12} showed that overlays and reference structures improve chart reading performance. Our conjoint results extend these single variable findings by showing that contextual scaffolding dominates preference even when traded off against other design dimensions (graphic type, color) simultaneously. This convergence across annotation studies, overlay experiments, and our conjoint data suggests a robust design principle: structural elements that aid calibration and comparison are valued more than purely aesthetic variation across a range of task contexts and evaluation methods.

\paragraph{Icon arrays and explicit encoding.}
Kandel et al.~\cite{KLX25} found that icon arrays with explicit encoding and juxtaposition optimize accuracy and decisions in health risk communication. The near equivalence of icon series (41.5\%) and bar charts (40.4\%) in our data mirrors their finding that both formats perform well when providing adequate comparative structure. The broader health risk literature~\cite{GGG09, Zik*14} consistently shows that explicit encodings and redundancy improve both comprehension and preference.

\paragraph{Integrated design agenda.}
A follow on study measuring preference, comprehension, trust, and image or information classification for the same stimuli would enable holistic guidance. The CBC format can accommodate additional ratings after each choice with minimal overhead, moving the field from preference only results toward convergent design evidence.

\section{Modern Conjoint Methods}
\label{app:modern_conjoint}

\paragraph{Hierarchical Bayes and mixed logit.}
HB estimation or mixed logit models estimate respondent level part worth distributions, identifying preference segments. With 65 respondents and 12 tasks each, HB is feasible but would yield wide individual level posteriors. Recent scalable variational Bayes (VB) mixed logit implementations~\cite{KBBD20} for panel data make this computationally lightweight even for larger follow ups: VB achieves up to 16 times speedup over MCMC with comparable parameter recovery, making it practical for visualization researchers unfamiliar with Bayesian estimation. Even aggregate level HB estimates would cross check the AMCE ordering.

\paragraph{Neural approaches.}
ConjointNet~\cite{ZCH25} uses representation learning to capture nonlinear interactions, outperforming linear models by over 5\% on benchmark datasets. For our data it could detect whether interactions (for example, bar by scale) contribute meaningfully, though our sample size risks overfitting.

\paragraph{Adaptive design.}
The Gradient based Survey (GBS)~\cite{YGLS23} adaptively constructs questions based on prior choices, requiring no parametric model. GBS is most valuable for large attribute spaces beyond our three factor design.

\paragraph{Subgroup discovery.}
Goplerud et al.~\cite{GIP22} proposed FactorHet, a mixture of experts framework that discovers latent respondent subgroups data adaptively while estimating heterogeneous effects. Unlike manual splits (for example, US and UK vs.\ others), FactorHet finds maximally heterogeneous groups using moderating covariates. Related BART based approaches (\texttt{cjbart}) detect individual level marginal component effects. ML based black box inference frameworks can further recover interaction effects without prespecification.

\section{Bradley Terry Scoring}
\label{app:bradley_terry}

The Bradley Terry (BT) model estimates latent ``strength'' scores from win and loss patterns. Fageot et al.~\cite{FFHV24} introduced a generalized BT (GBT) family with guaranteed unique MLE. BT scores complement AMCEs in two ways: (1) they are invariant to baseline level choice, producing a single ranking; (2) they provide a natural transitivity and internal consistency check; poor BT fit would indicate noisy or context dependent preferences, while good fit confirms coherent orderings. The raw data can be analyzed with \texttt{BradleyTerry2} (R) or \texttt{solidago} (Python).

\section{External Validity}
\label{app:external_validity}

Our single topic (unemployment at an extreme 50\%) and two color palette constrain generalization. The comparison type effect may be inflated for high stakes, personally relevant topics where contextual information is especially valued; for neutral topics (rainfall, historical populations), color and graphic type might gain relative importance. The 50\% rate is unrealistic, potentially making comparison information especially helpful. Red and blue are semantically loaded yet produced no difference; other colors (green, orange) might. Replication should test 2 to 3 additional topics spanning personal relevance and valence, a broader palette ($\geq 4$ colors), and realistic data values.

\paragraph{Crisis communication context.}
Schneider et al.~\cite{SKDA23} reviewed how format, context, comparisons, and visualization shape the communication of statistical evidence during crises. Their synthesis highlights that readers rely more heavily on contextual scaffolding (benchmarks, reference points, and calibrated scales) when stakes are high and information is rapidly changing. Our unemployment scenario, framed as a news report, aligns with this pattern: participants may have treated the task as a high stakes information seeking context where comparison cues are especially valued. This suggests the comparison type dominance we observe may be partly context driven: in low stakes or exploratory settings (for example, browsing data dashboards for entertainment), aesthetic attributes like color might carry more weight. Future replications should vary topic urgency to test this boundary condition.

\paragraph{Focal value sensitivity.}
The extreme 50\% unemployment rate may have amplified the comparison type effect in two ways. First, an extreme statistic increases the desire for context (``Is this really that bad?''), making benchmarks and scale markers especially valued. At more realistic values (for example, 6\%, 12\%, 25\%), the information need may be lower and comparison features less preferred. Second, the 50\% value happens to coincide with the midpoint of a 0 to 100\% axis, making the percentage scale condition particularly clean visually (a tick at 50\% aligns exactly with the data point). At other values (for example, 8.3\%), axis ticks would not align as neatly, potentially reducing the condition's visual advantage. A follow up varying the focal value across a realistic range (for example, 6\%, 12\%, 25\%, 50\%) would directly test whether the comparison type AMCE attenuates at lower magnitudes. If the effect shrinks substantially at realistic rates, the practical recommendation to ``always add context'' would need to be qualified as context dependent itself.

\section{Trial Order and Procedural Diagnostics}
\label{app:trial_order}

Each participant completed about 12 trials. Possible order effects include fatigue driven lexicographic simplification (inflating comparison type dominance in later trials), learning effects (noisier early trials), and position bias drift. Bansak et al.~\cite{Ban*21} found conjoint experiments reliable with many tasks, but their samples were larger.

If the Conjointly export includes trial sequence identifiers, three diagnostics are recommended: (1) \emph{early vs.\ late AMCE split} (trials 1 to 6 vs.\ 7 to 12); (2) \emph{choice consistency} on dominated pairs by trial position; (3) \emph{left and right position bias} by trial index. If per trial response times are available, flagging extremely fast responses (under 2 seconds) and checking for declining median response time would further assess engagement. We were unable to confirm whether trial order or response time data are available in the export; future studies should request these fields and preregister exclusion criteria (for example, under 1.5 second trials).

\section{Power Analysis}
\label{app:power}

The effective sample size under clustering is $N_{\mathrm{eff}} \approx N / (1 + (m-1)\rho)$. At $\rho = 0.05$, $N_{\mathrm{eff}} \approx 700$, allowing detection of about 7.5~pp main effects at 80\% power ($\alpha = 0.05$). Our comparison type ($+34.6$, $-41.6$~pp) and graphic type ($+18.1$, $-11.1$~pp) effects are well within range; color ($\pm 0.5$~pp) is far below, so the null result is consistent with either no effect or an undetectable one. Interaction effects require about 4 times the sample, yielding about 15~pp detectable; subtle interactions (5 to 10~pp) would be missed. A Monte Carlo simulation under observed AMCEs and varying $\rho$ could confirm these bounds more precisely; we recommend such simulation based power analysis \emph{before} data collection in future studies. Follow up targets: 300 to 400 respondents for 10~pp interactions, 800 plus for 5~pp or subgroup heterogeneity~\cite{Ban*21}.

\end{document}